\documentclass{PoS}
\usepackage{epsfig}
\usepackage{multicol}
\PoS{PoS(HEP2005)}

\def\beq{\begin{eqnarray}}
\def\eeq{\end{eqnarray}}

\def\be{\begin{equation}}
\def\ee{\end{equation}}

\def\slash#1{#1 \hskip-0.50em /}

\title{The $B \to \pi$ form factor from light-cone sum rules in soft-collinear effective theory\thanks{CERN-PH-TH/2005-167, SLAC-PUB-11484, BARI-TH/05-518, SI-HEP-2005-12}}

\ShortTitle{$B \to \pi$ form factor from LCSR in SCET}

\author{\speaker{Tobias Hurth}\thanks{Heisenberg Fellow}\\
        CERN, Dept.\ of Physics, Theory Unit, CH-1211 Geneva 23, Switzerland\\ SLAC, Stanford University, Stanford, CA 94309, USA\\
        E-mail: \email{tobias.hurth@cern.ch,hurth@slac.stanford.edu}}

\author{Fulvia De Fazio\\
         Istituto Nazionale di Fisica Nucleare, Sezione di Bari, Italy\\
    E-mail: \email{fulvia.defazio@ba.infn.it}}

\author{Thorsten Feldmann\\
Department of Physics, University of Siegen, D-57068 Siegen, Germany\\
   E-mail: \email{feldmann@hep.physik.uni-siegen.de}}
    
\abstract{Recently, we have derived
light-cone sum rules for exclusive $B$\/-meson decays 
into light energetic hadrons from correlation functions within soft-collinear effective theory \cite{ours}.
In these sum rules the
short-distance scale refers to ``hard-collinear'' interactions with
virtualities of order $\Lambda_{\rm QCD} m_b$. Hard scales (related
to virtualities of order $m_b^2$) are integrated out and enter via
external coefficient functions in the sum rule. Soft dynamics is
encoded in light-cone distribution amplitudes for the
$B$\/-meson, which describe both the factorizable and non-factorizable
contributions to exclusive $B$\/-meson decay amplitudes.
Factorization of the correlation function has been verified to
one-loop accuracy. Thus, a systematic separation of hard, hard-collinear, 
and soft dynamics in the heavy-quark limit is possible.}

\FullConference{HEP2005 International Europhysics 
                Conference on High Energy Physics\\
                21--27 July 2005\\
	        Lisbon, Portugal}

\begin{document}


 \newcommand{\bea}{\begin{eqnarray}}
 \newcommand{\eea}{\end{eqnarray}}
 \newcommand{\nn}{\nonumber}
 \newcommand{\dd}{\displaystyle}
 \newcommand{\bra}[1]{\left\langle #1 \right|}
 \newcommand{\ket}[1]{\left| #1 \right\rangle}
 \newcommand{\spur}[1]{\not\! #1 \,}


The QCD factorization theorems for exclusive, energetic $B$\/-decays,
first proposed in \cite{Beneke:1999br}, 
identify short-distance effects 
that can be systematically calculated in perturbation theory. 
Non-pertur- bative effects are parametrized in terms of a few universal 
functions such as  form factors and light-cone distribution amplitudes, 
on which our information is rather restricted. 
Moreover, phenomenological applications are limited by the
insufficient information on (power-suppressed) nonfactorizable terms. 
These restrictions should be tackled by the large data sets available by 
current and future $B$- physics experiments on the experimental side, and by the development of suitable non-perturbative methods on the theory side.

The quantum field theoretical framework corresponding to the 
QCD factorization theorems  is the soft collinear effective theory (SCET)
\cite{SCETa,SCETb}.  
 In contrast to the well-known 
heavy-quark effective theory (HQET), the recently 
proposed SCET does not correspond to a local operator expansion.
While HQET is only  applicable to $B$ decays, when the energy transfer 
to light hadrons  is small, for example to  $B \rightarrow D$ transitions 
at small recoil to the $D$ meson, it is not applicable, when 
 some of the outgoing, light particles have momenta of order $m_b$; 
then one faces a multiscale problem:
a) $\Lambda = {\rm few} \times \Lambda_{\rm QCD}$,
the {\it soft} scale set by the typical energies and
momenta of the light degrees of freedom in the hadronic bound states;\,\,
b) $m_b$, the {\it hard}\/ scale set by the heavy-$b$\/-quark mass;\,\,
c) the hard-collinear scale $\mu_{\rm hc}=\sqrt{m_b
\Lambda}$ appears via interactions between soft and energetic
modes in the initial and final states. The dynamics of hard and hard-collinear
modes can be described perturbatively in the heavy-quark limit 
$m_b \to \infty$.
The separation of the two perturbative scales 
from the non-perturbative hadronic dynamics is determined
by the small expansion parameter 
$\lambda = \sqrt{\Lambda/m_b}$.
On a technical level the implementation of power counting 
in $\lambda$ for fields and operators in SCET corresponds 
directly to the well-known method of regions for Feynman diagrams \cite{Regions}.

A simple example where the above considerations apply is
given by the $B \to \pi$ transitions form factor.
In the large recoil-energy limit  the heavy-to-light 
form factors obey  relations  \cite{Charles:1998dr} 
that are broken by radiative and power corrections.
Each form factor can be decomposed into two basic contributions,
one piece that factorizes into a perturbatively calculable
coefficient function $T_i$ and light-cone distribution amplitudes
$\phi_B$ and $\phi_\pi$
for heavy and light mesons, respectively, and a  second contribution, 
where the hard-collinear interactions are
{\em not}\/ factorizable, leaving one universal 
``soft'' form factor $\xi_{\pi}$ \cite{Beneke:2000wa}:
$$
\langle \pi| \bar \psi \, \Gamma_i \, b |B\rangle
=  C_i(E, \mu_I) \, \xi_\pi(\mu_I,E) +
   T_i(E,u,\omega,\mu_{\rm II}) \otimes \phi_+^B(\omega,\mu_{\rm II})
   \otimes
         \phi_\pi(u,\mu_{\rm II})+
   \mbox{subleading terms}\,. \label{factorization}
$$
Here $C_i$ is a short-distance function arising from integrating out
hard modes, and $T_i$ contains hard and hard-collinear dynamics related to spectator scattering.
Consequently $\mu_I$ is a factorization scale
below $m_b$, while $\mu_{\rm II}$ is 
a factorization scale below $\mu_{\rm hc}$.
Both functions can be computed as perturbative series in $\alpha_s$
(the effective theories for the two short-distance 
regimes are known as SCET$_{\rm I}$ and SCET$_{\rm II}$).
We have shown in \cite{ours} that light-cone sum rules can be formulated for 
the {\em soft}\/ (i.e.\ non-factorizable)
part of the form factor {\em within} SCET$_{\rm I}$. 
In the following we briefly discuss the basic idea using 
the tree-level construction in an examplary mode.
For the derivation of all further non-trivial results, the 
phenomenological application, and also for the general notation used
in this short letter,  
we guide the reader to the original 
publication \cite{ours}. 

Within SCET$_{\rm I}$
the non-factorizable (i.e.\ end-point-sensitive) part
of the $B \to \pi$ form factor in
the heavy-quark limit is described by the current operator
\beq
 J_0(0) = \bar \xi_{\rm hc}(0) W_{\rm hc}(0) Y_s^\dagger(0) h_v(0)\,\,,
\qquad
\langle \pi(p') |\, J_0(0)\,|B(m_B v)\rangle
  =
  (n_+p')\,{\xi_\pi(n_+p', \mu_{\rm I})}\,, 
\eeq
where $\xi_{\rm hc}$ is the ``good'' light-cone component of the
light-quark spinor with $\slash n_- \xi_{\rm hc} =0$, 
and $W_{\rm hc}$ and $Y_s^\dagger$
are hard-collinear and soft Wilson lines. Finally, $h_v$
is the usual HQET field. 
The heavy quark is nearly on-shell in the end-point
region. In SCET$_{\rm I}$, this is reflected by
the fact that hard sub-processes (virtualities of order $m_b^2$) are already
integrated out and appear in coefficient 
functions multiplying $J_0$, which can be determined from the matching of the
corresponding QCD current on SCET$_{\rm I}$.
Therefore we will {\em not}\/ introduce 
an interpolating current for the $B$ meson as in the usual light-cone 
sum-rule approach. 
Instead, the short-distance
(off-shell) modes in SCET$_{\rm I}$ are the hard-collinear
quark and gluon fields, and therefore 
the sum rules should be derived from a dispersive
analysis of the correlation function
\beq
 \Pi(p') = i \int d^4x \,
   e^{i p'{} x} \, \langle 0 | T[ J_\pi(x) J_0(0)] | B(p_B) \rangle\,, 
\eeq
where $p_B^\mu = m_B v^\mu$, and the interpolating current $J_{\pi}$ 
for a pion in the effective theory 
is chosen as
\beq
 J_\pi(x) 
= - i \, \bar \xi_{\rm hc}(x)  \, 
          \slash n_+ \gamma_5 \, \xi_{\rm hc}(x) - i \,
\left( \bar \xi_{\rm hc} W_{\rm hc}(x)  
       \slash n_+ \gamma_5 Y_s^\dagger q_s(x) + h.c.\right) \,.
\eeq
with $\langle 0| J_\pi | \pi(p')\rangle = (n_+p') \, f_\pi$.
Here we denoted 
soft and hard-collinear quark fields in SCET
as $q_s$ and $\xi_{\rm hc}$, respectively.
Notice that soft-collinear interactions require a multipole expansion
of soft fields, which is always understood implicitly. 
We also point out that the effective theory SCET$_{\rm I}$
contains SCET$_{\rm II}$ as its infrared limit (i.e.\ when
the virtuality of the hard-collinear modes is lowered to 
order $\Lambda^2$). For this reason, the hard-collinear fields
that define the interpolating current $J_\pi$ also contain the
collinear configurations that show up as hadronic bound states
(see also the discussion in \cite{Beneke:2003pa}).

\begin{figure}[tbph]
\begin{center}
\epsfig{file=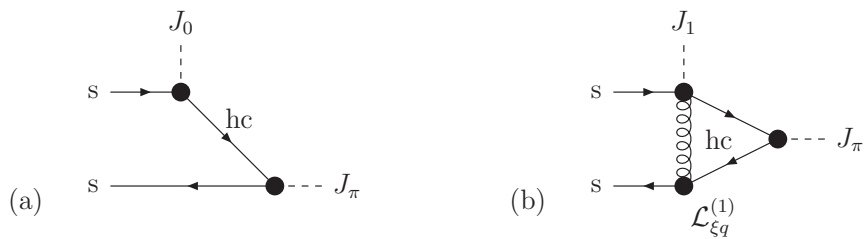, width=0.75\textwidth}
\end{center}\vspace{-0.8cm}
\caption{(a) Leading contribution to the correlation function for the
non-factorizable SCET current $J_0$.\,\,  (b) Leading contribution to the
sum rule for the factorizable SCET current $J_1$.}
\label{fig:lead}
\end{figure}

In the following we will consider a reference frame where
$p'_\perp=v_\perp=0$ and $n_+v = n_-v = 1$.
In this frame the two independent kinematic variables are
$(n_+ p'{}) \simeq 2 E_\pi={\cal O}(m_b)\, \mbox{and}\,  
 0>(n_- p'{})={\cal O}(\Lambda)$
with $|n_-p'| \gg m_\pi^2/(n_+p')$.
The dispersive analysis will be performed with respect to
$(n_-p')$ for fixed values of $(n_+p')$.
As with all QCD sum-rule calculations, the procedure consists 
in writing the correlator in two different ways. 
On the hadronic side, one can write 
\beq
\Pi^{\rm HAD}(n_-p')= \Pi(n_- p') \Big|_{\rm res.}+ \Pi(n_- p')
\Big|_{\rm cont.} \,,
\eeq
where the first term represents the contribution of the pion, while the second takes into account
the role of higher states and continuum above an effective
threshold $\omega_s = {\cal O}(\Lambda^2/n_+p')$. The former can be rewritten
as 
\beq
 \Pi(n_- p') \Big|_{\rm res.} = 
   \frac{\langle 0|J_\pi |\pi(p')\rangle \langle 
   \pi(p')|J_0|B(p_B)\rangle}
    {m_\pi^2-p'{}^2}
 \ = \
 \frac{ (n_+p')^2 \, \xi_\pi(n_+p') \, f_\pi}{m_\pi^2-p'{}^2}\,.
\eeq
On the SCET side of the sum rule, the tree-level result (see Fig.~\ref{fig:lead}(a)),  
for the correlation function involves one hard-collinear quark propagator,
which reads
\beq
  S_F^{\rm hc} = \frac{i}{n_- p'{} - \omega + i\eta}
  \, \frac{\slash n_-}{2}\,,
\eeq
where $\omega = n_- k$, and $k^\mu$ is the momentum of the soft light quark that will end up as the spectator quark in the $B$ meson.
The propagator is always off shell and always induces light-like separations as long as $|n_- p'{}| \sim \Lambda$.
This leads to  matrix elements of operators that are formulated
only in terms of
soft fields, which are separated along the light cone and thus define
light-cone distribution amplitudes for the $B$\/ meson in HQET
\cite{Grozin:1996pq}.
The final result  already has the form of a dispersion integral in the variable $n_- p'$ : 
\beq
  \Pi(n_- p'{}) = 
  f_B m_B \int_0^\infty d\omega \,
   \frac{\phi_-^B(\omega)}{\omega - n_-p'{}-i\eta} \,,
\eeq
where the $B$ light-cone distribution amplitude enters through 
\beq
 \langle 0|\bar q(x_-) \, Y_s\, \gamma_5 \, \frac{\slash n_+ \slash 
 n_-}{2} \, Y_s^+  \, h_v(0)|B(p_B)\rangle =
  i f_B \, m_B  \,
   \int  d\omega' \, e^{-i\omega' \, \frac{n_+ x}{2}} \,
   \phi_-^B(\omega')\,.
\eeq
The result for the SCET side of the sum rule shows 
that the considered correlation 
function in the (unphysical)
Euclidean region factorizes into a perturbatively
calculable hard-collinear kernel and a soft light-cone
wave function for the $B$ meson, where the convolution variable
$\omega$ represents the light-cone momentum of the spectator quark
in the $B$ meson.

Finally, $\Pi(n_- p') \Big|_{\rm cont.}$ on the hadronic side 
can be written again according to a dispersion
relation. Moreover,  assuming global
quark--hadron duality, we identify the spectral density with its
perturbative expression above some threshold $\omega_s$.
The Borel transform with respect to the variable $n_-p'{}$ introduces
the Borel parameter $\omega_M$ and reads in this case 
$\hat B (\omega_M) [\,1 / (\omega-n_-p')\,]  = 1 / \omega_M \,\,  e^{-\omega/\omega_M}$.
As usual, the physical role of the Borel parameter is to enhance the
contribution of the hadronic-resonance region, where the virtualities
of internal propagators have to be smaller than the hard-collinear scale. 
Equating the two representations of the correlator, using
global quark--hadron duality (and neglecting the pion mass), we obtain  
the final sum rule for the soft form factor at tree level 
(see also \cite{Khodjamirian:2005ea}): 
\beq
 \xi_\pi(n_+p') =  \frac{f_B m_B}{f_\pi (n_+p')} \,
   \int_0^{\omega_s} d\omega \, e^{-\omega/\omega_M}  \,
   \phi_-^B(\omega)\,.
\eeq
The result for $\xi_\pi$ has the correct scaling with
$\Lambda/m_B$ as obtained from the conventional sum rules or
from the power counting in SCET. 
The resulting estimate for
the soft form factor at maximum recoil, $0.27^{+0.09}_{-0.11}$, is compatible
with other determinations. The uncertainty is dominated by the
variation of the sum-rule parameters and by the product
$f_B \phi_-^B$.

In \cite{ours}, we have also reproduced the result for the factorizable form-factor contribution from 
hard-collinear spectator scattering
\cite{Beneke:2000wa}. For this purpose we have  to consider 
the correlation function
\beq
  \Pi_1(p') = i \int d^4x \,
   e^{i p'{} x} \, \langle 0 | T[ J_\pi(x) J_1(0)] | B(p_B) \rangle \,,
\quad  J_1 \equiv
 \bar \xi_{\rm hc} \,  g \, \slash A^\perp_{\rm hc} \,  h_v\,,
\eeq
where we used the light-cone gauge. 
The leading contribution is given by the diagram in Fig.~\ref{fig:lead}(b)
which involves the insertion of one interaction vertex from
the order-$\lambda$ soft-collinear Lagrangian.
We derived the remarkable feature of the SCET-sum-rule approach to the $B \to
\pi$ form factor that the ratio of factorizable and
non-factorizable contributions is independent of the $B$-meson
wave function to first approximation, and about $6\%$, 
which is in line with the power counting used in QCD factorization
\cite{Beneke:1999br}, but contradicts the assumptions
of the so-called pQCD approach \cite{Chen:2001pr}.
The non-trivial issues related to the factorization of
the correlation function arise beyond tree level. 
We have shown that the ${\cal O}(\alpha_s)$ 
short-distance radiative corrections from hard-collinear 
loops preserve factorization, i.e.\ the encountered IR divergences 
correspond to the renormalization of the $B$\/-meson distribution
amplitude(s) in HQET (three-particle Fock states
have been neglected so far). At this point our approach differs
from the conventional sum rules formulated in QCD,  where the
separation of three different scales ($m_b$, $\mu_{\rm hc}$,
$\Lambda$) is usually not attempted. 
However, in a recent article \cite{Lee}
Lee proposes to formulate sum rules within the effective field 
theory framework of SCET for the conventional
set-up where the pion is represented by a 
collinear light-cone distribution amplitude and the $B$ meson
is interpolated by a current in HQET. We stress that in that
case the radiative corrections would involve soft {\em and}\/ 
collinear momentum regions; according to the general 
discussion in \cite{Beneke:2003pa,Lange:2003pk,Bauer:2002aj}, 
 these are not expected to factorize. Indeed the occurrence 
of end-point singularities that spoil factorization has been 
observed in \cite{Lee}.
Our choice of interpolating the pion instead avoids the
problem of collinear end-point divergences. In this way the 
SCET-based sum rules generalize the ideas of QCD factorization 
to the non-factorizable parts of exclusive decay amplitudes, 
and allow for a systematic and consistent expansion in terms of 
$1/m_b$ and $\alpha_s$, where the hadronic information is encoded 
in light-cone distribution amplitudes for the $B$ meson and 
non-perturbative sum-rule parameters. At present, 
 because of the limited information on the $B$\/-meson wave function, the theoretical uncertainties are larger than those found in QCD light-cone sum rules. 

\begin{multicols}{2}

\end{multicols}

\end{document}